\documentclass[conference]{IEEEtran}
\IEEEoverridecommandlockouts
\usepackage{cite}
\usepackage{amsmath,amssymb,amsfonts}
\usepackage{algorithmic}
\usepackage{graphicx}
\usepackage{textcomp}
\usepackage{xcolor}

\usepackage[T1]{fontenc}
\usepackage[utf8]{inputenc}
\usepackage{authblk}
\usepackage{diagbox} 

\usepackage[figuresright]{rotating}
\usepackage{subfigure}
\usepackage{wrapfig}
\usepackage{url}
\usepackage{epsfig}
\usepackage{color}
\usepackage{epstopdf}
\usepackage[ruled]{algorithm2e} 

\SetAlFnt{\small}
\SetAlCapFnt{\small}
\SetAlCapNameFnt{\small}
\SetAlCapHSkip{0pt}
\IncMargin{-\parindent}

\def\BibTeX{{\rm B\kern-.05em{\sc i\kern-.025em b}\kern-.08em
    T\kern-.1667em\lower.7ex\hbox{E}\kern-.125emX}}
\begin{document}


\title{A Container-based DoS Attack-Resilient Control Framework for Real-Time UAV Systems}

\author[1]{Jiyang Chen}
\author[2,1]{Zhiwei Feng}
\author[1]{Jen-Yang Wen}
\author[3]{Bo Liu\textsuperscript{*} \thanks{\textsuperscript{*}Bo Liu is now with {NVIDIA} Corporation, USA}}
\author[1]{Lui Sha}
\affil[1]{Department of Computer Science, University of Illinois at Urbana-Champaign, USA}
\affil[2]{School of Computer Science and Engineering, Northeastern University, China}
\affil[3]{{Coordinated Science Laboratory, University of Illinois at Urbana-Champaign, USA}}
\maketitle

\begin{abstract}
The Unmanned aerial vehicles (UAVs) sector is fast-expanding.
Protection of real-time UAV applications against malicious attacks has become an urgent problem that needs to be solved. 
Denial-of-service (DoS) attack aims to exhaust system resources and cause important tasks to miss deadlines. DoS attack may be one of the common problems of UAV systems, due to its simple implementation.  
In this paper, we present a software framework that offers DoS attack-resilient control for real-time UAV systems using containers: ContainerDrone.
The framework provides defense mechanisms for three critical system resources: CPU, memory, and communication channel. 
We restrict attacker's access to CPU core set and utilization.
Memory bandwidth throttling limits attacker's memory usage. 
By simulating sensors and drivers in the container, a security monitor constantly checks DoS attacks over communication channels. 
Upon the detection of a security rule violation, the framework switches to the safety controller to mitigate the attack. 
We implemented a prototype quadcopter with commercially off-the-shelf (COTS) hardware and open-source software.
Our experimental results demonstrated the effectiveness of the proposed framework defending against various DoS attacks.


\end{abstract}

\begin{IEEEkeywords}
Cyber Physical System, Real-time System, Denial of Service attack, Linux Container, Simplex, Unmanned Aerial Vehicle Systems, Security
\end{IEEEkeywords}
\section{Introduction}\label{s:intro}


    


Recent advancement of Unmanned Aerial Vehicle (UAV) technologies has promised a booming industry that brings novel UAV applications in civil, commercial, and military sectors \cite{valavanis2015future}. As a great example of a cyber-physical system (CPS), modern UAVs are networked robots equipped with capable communication channels, sophisticated sensor system and software, which offers advanced functionalities, however, it also exposes the UAV system to malicious attacks. With UAVs being able to carry out more tasks, reported attacks targeting UAVs \cite{hijack_mavlink, kerns2014unmanned} have raised concerns about safety and security of the UAVs due to their increased attacking surface. A successful attack on UAVs may lead to disastrous property damage or even loss of life. Guarding UAV systems against malicious attacks is of great significance.

The Denial-of-service (DoS) attack which slows the exchange of information by sending superfluous requests has become a significant problem for Internet communication \cite{senie1998network}.
It aims to deny the access of legitimate users to shared services or resources to make a machine or network resource unavailable \cite{gligor1984note}, which can cause system overload and prevent some or all legitimate requests from being fulfilled.
Even a temporary DoS attack can compromise the strong real-time constraints of many unmanned control systems.
Though some UAVs might not be connected to the Internet, they might still be the victim of the DoS attack as malicious code might be hidden inside installed software seeking opportunities to attack. 
Defending against internal DoS attack requires constant monitoring of all processes running on the system and their resource usage, which can add extra overhead to the system. 
Using more powerful hardware can alleviate this issue, but the compromise made in size, weight, and power consumption makes this solution unsuitable for embedded real-time UAVs.


To address this security challenge and meet the real-time requirement of UAVs, we present the ContainerDrone framework, a container-based software framework that offers DoS attack-resilient control for real-time UAVs. 
It provides two distinct control environments: a verified host control environment that guarantees safety but offers only basic functionalities, a potentially vulnerable container control environment that provides optimized performance and advanced functionalities. 
We protect three critical system resources: CPU, memory, communication channel, using Linux cgroup, Docker \cite{dockerintro} built-in mechanism, MemGuard \cite{yun2013memguard}, and security monitoring.  


Our paper makes the following contributions:
\begin{itemize}
    \item We developed a novel DoS attack-resilient control framework, ContainerDrone, for real-time UAV system using containers. The host control environment is protected against DoS attack from three perspectives: CPU, memory, and communication.  
    \item We implemented the framework on a quadcopter with an off-the-shelf multi-core embedded board and an autopilot sensor board. The operating system and software used in building this framework are all from open-source projects.
    \item We performed experiments to show the capabilities of the framework and how different kinds of DoS attack can be defended. 
\end{itemize}

The structure of this paper is organized as the following:
Section \ref{s:bg} introduces the background knowledge of Docker container and Simplex architecture.
The detailed idea of the proposed framework is discussed in Section \ref{s:framework}.
In Section \ref{s:imple}, we describe the implementation of our prototype drone using the proposed framework. 
We presents our experiments and results in Section \ref{s:exp}.
The related work is discussed in Section \ref{s:rela} 
and finally Section \ref{s:conclusion} summarizes this work.


\section{Background and related work}\label{s:bg}

\subsection{Containers}


Container is a popular open-source technology that provides lightweight software isolation \cite{merkel2014docker,preeth2015evaluation}. 
In its core, it is an abstraction of several features in Linux kernel. 
Control Groups (cgroups) provides resource allocation, such as CPU and memory. 
Namespace isolates processes so each has a different view of the system. 
Other modules like SELinux \cite{smalley2001implementing}, AppArmor \cite{apparmor}, Seccomp \cite{seccomp} provide container security capabilities \cite{bui2015analysis}. 
Instead of virtualizing an entire operating system, containers provide only the bare minimum requirements for the application to run and multiple containers share the host operating system kernel. On the host system, each container is just a process running the application and necessary dependencies. 
System isolation using containers leads to less execution overhead, less memory usage, and reduced footprint.

Compared with virtual machines, containers have several advantages \cite{dockerintro}:
\begin{itemize}
    \item Lightweight: Containers share the host's kernel while the virtual machine needs a hypervisor to talk to the kernel and the underlying hardware. 
    \item Efficiency: The average size of a container is around tens of MB while a VM easily exceeds GB. More containers can run on the same machine at the same time than VMs could which means the same computing resource may generate more results using containers. 
    \item Portability: Containers have developed over the years and evolved from a Linux only application to be able to run virtually anywhere: on Mac and Windows operating system; on bare metal and public cloud. It greatly reduces the effort for development and time for deployment. 
\end{itemize}

\subsection{Simplex Architecture}\label{s:simplex}

    

\begin{figure}[h]
	\centering
	\includegraphics[width=0.7\linewidth]{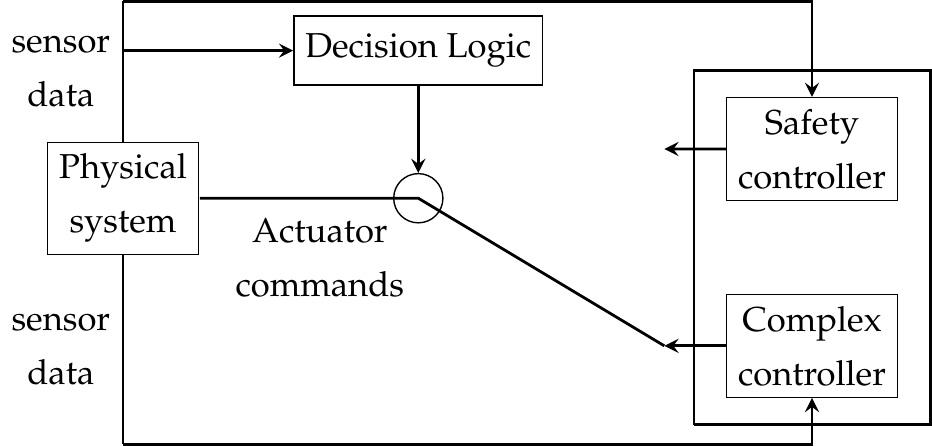}
	\caption{The control and switching logic of the Simplex architecture}\label{f:simplex}
\end{figure}

The Simplex architecture\cite{sha2001using} is a software approach that provides a reliable control system.
Instead of trying to avoid all possible faults, the core idea of Simplex is fault tolerance. 
As shown in figure \ref{f:simplex}, it is composed of a complex controller, a safety controller, and a decision module. 
The complex controller implements advanced functionalities and provides optimized performance, but is potentially unverifiable due to its complexity. 
The safety controller has only limited performance but is robust thanks to its simplicity. 
It can be exhaustively tested and verified to be safe. 
During runtime, all necessary sensor data are sent to both controller and each of the controllers computes their own actuator outputs. 
The decision module uses a safety envelop to determine if any safety violation occurs, and is responsible for deciding which output should be used for the current situation. 
Under normal execution, the outputs from the complex controller are used as it can provide better performance.
In the case of a fault or attack, the decision module detects the violation and switches to the safety controller to bring the system back to a safe and controllable state. 





\section{ContainerDrone framework}\label{s:framework}

The ContainerDrone provides defense against DoS attack by protecting three critical resources: CPU, memory, and communication channel. The Simplex architecture is implemented to provide attack-resilience. 


\begin{figure}[t]
	\centering
	\includegraphics[width=0.7\linewidth]{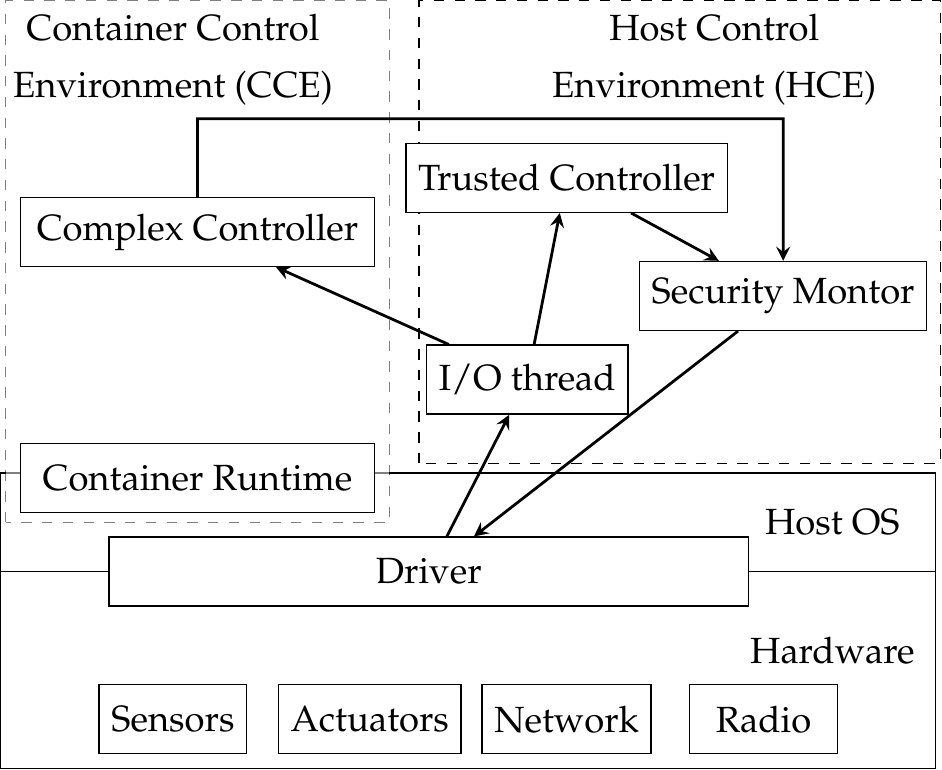}
	\caption{Simplex architecture}\label{f:framework}
	\vspace{-1em}
\end{figure}

\subsection{System model}
The system is composed of two separate control environments that offer different levels of performance and security protection. 

As the figure \ref{f:framework} shows, the container control environment (CCE) corresponds to the container environment and the complex controller inside it. 
The complex controller provides optimized performance and advanced features like mission planning and object avoidance. 
The software running inside the CCE are usually provided by third-party developers, have complex architectures and require constant updates. 
Their features make it not practical to verify them and leave potential vulnerabilities to attackers.

The host control environment (HCE) refers to the host OS and the safety controller running on top of it. 
It is designed for safety propose and should follow a simple design approach such that they can be fully analyzed and proved. 
Advanced features like collision avoidance will not be 
We assume the host OS is a simple-structured RTOS that its reliability has been verified, such as the seL4 micro-kernel \cite{klein2009sel4}. 
We assume all system processes running inside host OS is safe.
The safety controller runs a minimum set of modules that are critical to the correct functioning of the UAVs. 
It can take over the control from the CCE when the complex controller fails due to a DoS attack and thus guarantee the safety of the drone. 
A safety monitor runs on the host control environment and keeps monitoring the output from both controllers. 
Upon the detection of a security violation, the monitor changes the output source from the complex controller to the safety controller to prevent further damage. 

\subsection{Attacker model}

Applications in CCE may be safe during initial setup, but the attacker can embed malicious code in the unverified complex controller and get inside through updates. 
Inside the CCE, the adversary can launch a DoS attack against HCE by running any necessary program. 
However, the attacker does not have the capability to escape from the container.
According to \cite{de2017survey, bui2015analysis}, Docker is secure in terms of isolation. 
We trust the isolation provided by Docker as well as the hardware platform and the HCE.
The goal of the attacker is to use DoS attack to affect the host control environment and make the drone crash. 
In this paper, we do not consider physical component failures, software failure caused by bugs and logical faults, or any attacks other than DoS attacks. 

\subsection{CPU DoS Protection}
For the protection of CPU resource, we utilize the Linux kernel feature control group (cgroup) and Docker's built-in mechanism. We restrict CCE's access to CPU from two perspectives:
\begin{itemize}
    \item CPU core number: Cgroup's cpuset can bind the CCE to a set of CPU cores. Docker will also limit all its child processes to run on the assigned cpuset only. 
    \item CPU utilization: Docker restricts the process's ability to raise their priority. When the real-time First in-first out (FIFO) scheduling policies is used, a lower priority process can only get CPU share when a high priority job has finished. This prevents the processes in CCE to steal CPU share from HCE.  
\end{itemize}


\subsection{Memory DoS protection}

Although cgroup and Docker can limit memory node, and memory size a container can use, a malicious application in the container can still launch the DoS attack on the memory bandwidth by intensive accessing a small amount of main memory, as later demonstrated in the experiments. 

We use MemGuard~\cite{yun2013memguard} to protect our system from the memory bandwidth DoS attack.
MemGuard is a Linux kernel module that ensures each CPU core in the system does not access the memory exceeding a certain rate.
It uses the performance counter provided by the hardware to detect the number of memory access from each core for each MemGuard period.
Within a period, if the number of memory access initiated from a core exceeds the budget, MemGuard will restrict memory access from this core until the budget is replenished in the next period.
It has been used on DeepPicar \cite{bechtel2017deeppicar} to solve resource contention issue. 

However, there are other resource monitoring and bounding techniques, such as  \cite{inam2014multi,flodin2014dynamic}, which can also be used for memory DoS protection. 

\subsection{Communication DoS protection}

The complex controller inside CCE needs sensor data and user input commands to function correctly. However, it's also important to keep a strong separation between them for safety purpose. This is achieved by requiring simulation control mode in CCE and using security monitoring in HCE.

\textbf{Simulation Control Mode: }
Sensors and actuators in HCE are critical components that need to be protected from the untrustworthy software running in CCE. 
We require the complex controller to run in a simulation mode, where it does not access any device file but receive all the necessary data from the network interface.
Feeder threads running in HCE receives raw sensor data from device drivers and send them to both controllers. 
This eliminates all possible sensor DoS attack as no sensor device files exist in CCE. 

\textbf{Security Monitoring: }
The complex controller uses the network interface to communicate with HCE.
To protect the HCE from potential network DoS attack, the network stack for the two control environments are separated. 
The CCE lives in a sandboxed network space where it does not have access to the Internet and can only communicate with the HCE through a specified interface. 
Iptables is used to limit communication package rate of the network interfaces to reduce damage caused by DoS attacks.
A security monitor keeps monitoring the outputs received from the interface and also the physical state of the drone. 
Two security rules are enforced and upon a violation, the monitor kills the receiving thread on the HCE and switches to use the output from the safety controller. 
\begin{itemize}
    \item Receiving interval: The interval between two consecutive output received by the HCE should not be longer than a threshold. A long interval suggests the complex controller may have failed. 
    \item Attitude errors: The attitude (i.e., roll, pitch, and yaw) errors should be bounded at all time. The attitude errors reflect the drone's physical state. Large errors suggest the drone is in a dangerous state and might crash. 
\end{itemize}

\section{Implementation}\label{s:imple}

\subsection{Hardware platform}
\begin{figure}
\centering\
\subfigure[Prototype drone]{
		\includegraphics[width=0.5\linewidth]{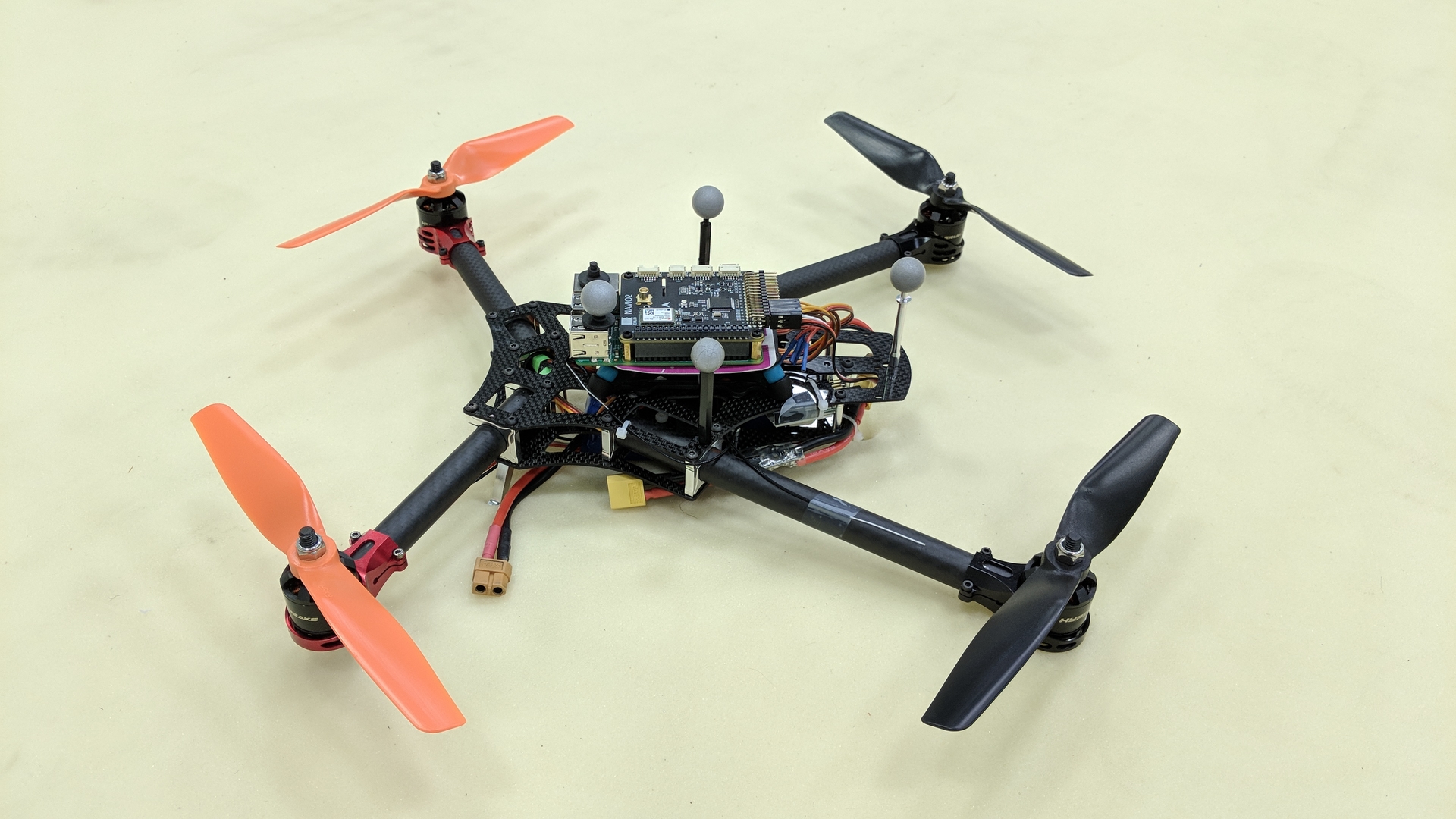}}
\subfigure[Navio2]{
		\includegraphics[width=0.3\linewidth]{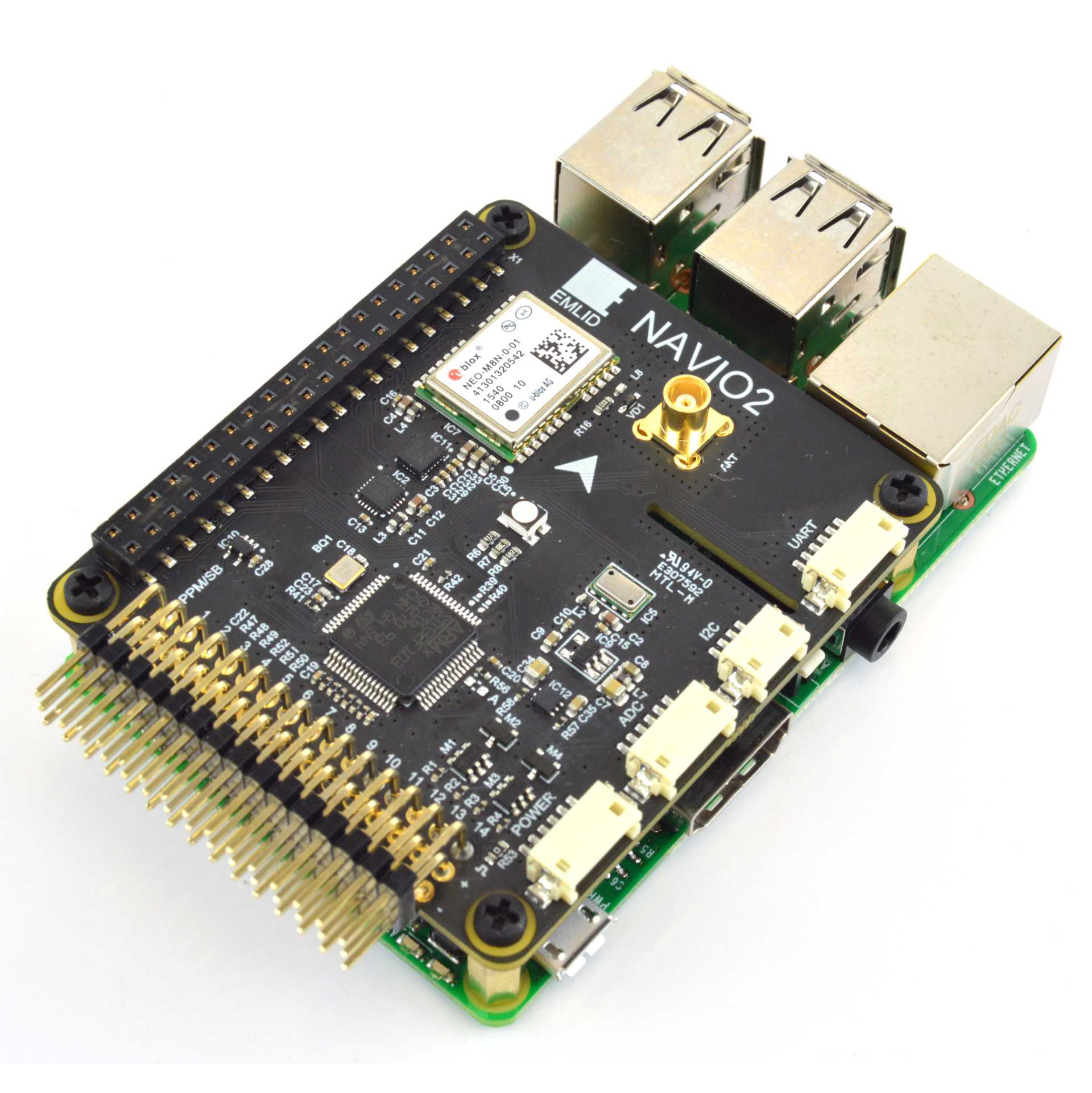}}
\caption{Prototype drone implemented with a Raspberry Pi 3 computer and a Navio2 sensor board}\label{f:hw}
\end{figure}

The ContainerDrone is implemented on a prototype quadcopter powered by a Raspberry Pi 3 Model B computer (RPi3B for short) as shown in Figure \ref{f:hw}.
It is equipped with a quad-core ARM Cortex A53 CPU running at a default frequency of 1.2GHz and has 1 GB of LPDDR2-900 SDRAM. 
Its small size and lightweight make it a excellent fit for a UAV application. 
On top of the RPi3B, we stack the Navio2 sensor hat from Emlid~\cite{EMLID}. It has two IMU chips, barometer and GNSS receiver, and provides sensing for RPi3B to control the flight control. 

\subsection{Operating System}
On RPi3B we use Linux 4.4\footnote{
There is no existing autopilot hardware that work with any verified OS, we use Linux as a proof of concept.
}
for the HCE. 
The kernel has been real-time patched and integrated with driver for Navio2.
On top of it, we run the container software Docker, a mature open source container platform, for CEE.
Docker wraps everything an application need, code, runtime, library, etc. in a package called container image. 
In the container, we run a Raspbian Jessie based Docker image provided by Resin.io, available from Docker Hub. 
One of the four cores is assigned exclusively for CCE use. 

We use Docker's port mapping to expose container ports to host OS. 
Hairpin NAT is enabled so port mapping is achieved only through modification of iptables rules, no port binding or user proxy process is involved.
This allows the host to communicate with container through defined UDP ports. 
No privilege flags are used in creating the container for CCE. 

We choose Linux as the host OS and Raspbian Jessie as the container image to minimize developing effort. 
It's recommended to use a verified OS for the host for security concerns. 
As the framework runs in userspace, no kernel level modification is required for neither host OS or container image. 

\subsection{Autopilot}
We use the PX4 \cite{px4dev} as the autopilot for both HCE and CCE. 
PX4 is an open source autopilot first developed at ETH Zurich and has now evolved into a large community project. 
PX4 offers one code base for all type of vehicles, including multi-copter, fixed-wing, and even ground vehicles. 
It supports MAVLink, a very lightweight robotic messaging protocol, for communication between Ground Control Station(GCS) and onboard components.  PX4 also uses a modular design. This makes it easy to integrate additional functionality into the existing code base. 

Two kernel driver processes used by PX4 on HCE are assigned with 90 FIFO scheduling priority while the safety controller on HCE is assigned 20 FIFO priority. These priority are chosen such that the kernel driver has the highest priority in the system and the safety controller has a priority lower than system interrupts (about 40 priority, assigned by Linux) but higher than the other processes.

\subsection{Communication between HCE and CCE}


The HCE and CCE communicate through a network interface. The CCE is configured to use a user-defined network where it does not have Internet access and can only access HCE through a dedicated docker0 interface. 
Sensor drivers in the HCE forward new data to complex controller through a UDP socket following the Mavlink protocol at the fixed sampling rate.
After the complex controller produces the actuator output, it is forwarded to HCE using UDP socket.
The rate and amount of data transfer from HCE and CCE and their receiving ports are summarized in Table \ref{t:com}.

\begin{table}[htbp]
\caption{The rate and amount of data transfer between the reliable and normal control environments.}
\begin{center}\label{t:com}
\begin{tabular}{|c|c|c|c|c|}
\hline
Component & Direction & Rate & Size & Port \\
\hline
IMU & HCE $\rightarrow$ CCE & 250Hz & 52 bytes & 14660 \\
\hline
Barometer & HCE $\rightarrow$ CCE & 50Hz & 32 bytes& 14660 \\
\hline
GPS & HCE $\rightarrow$ CCE & 10Hz & 44 bytes& 14660 \\
\hline
RC & HCE $\rightarrow$ CCE & 50Hz &  50 bytes & 14660\\
\hline
Motor Output & CCE $\rightarrow$ HCE & 400Hz & 29 bytes& 14600 \\

\hline
\end{tabular}
\label{tab1}
\end{center}
\end{table}



\section{Experiments}\label{s:exp}
In this section, we present the experiments that demonstrated the effectiveness of container-based framework against different types of DoS attacks. All experiments are conducted at the Intelligent Robotics Laboratory at the University of Illinois, Urbana-Champaign. A Vicon motion capture system is used to provide indoor positioning~\cite{liu2018viconmavlink} for the prototype drone such that it can fly autonomously in a controlled environment. In all the experiments, the drone operator first flies the drone to a safe height in manual mode and then switches to position control mode, where the drone uses all sensor and localization data to stabilize itself at a 3D set point in the space. The drone's Cartesian trajectory data along X, Y and Z axes are used for analysis. 

\begin{figure*}[t]
	\begin{minipage}[t]{0.23\linewidth}
		\centering
\includegraphics[width=1.0\linewidth]{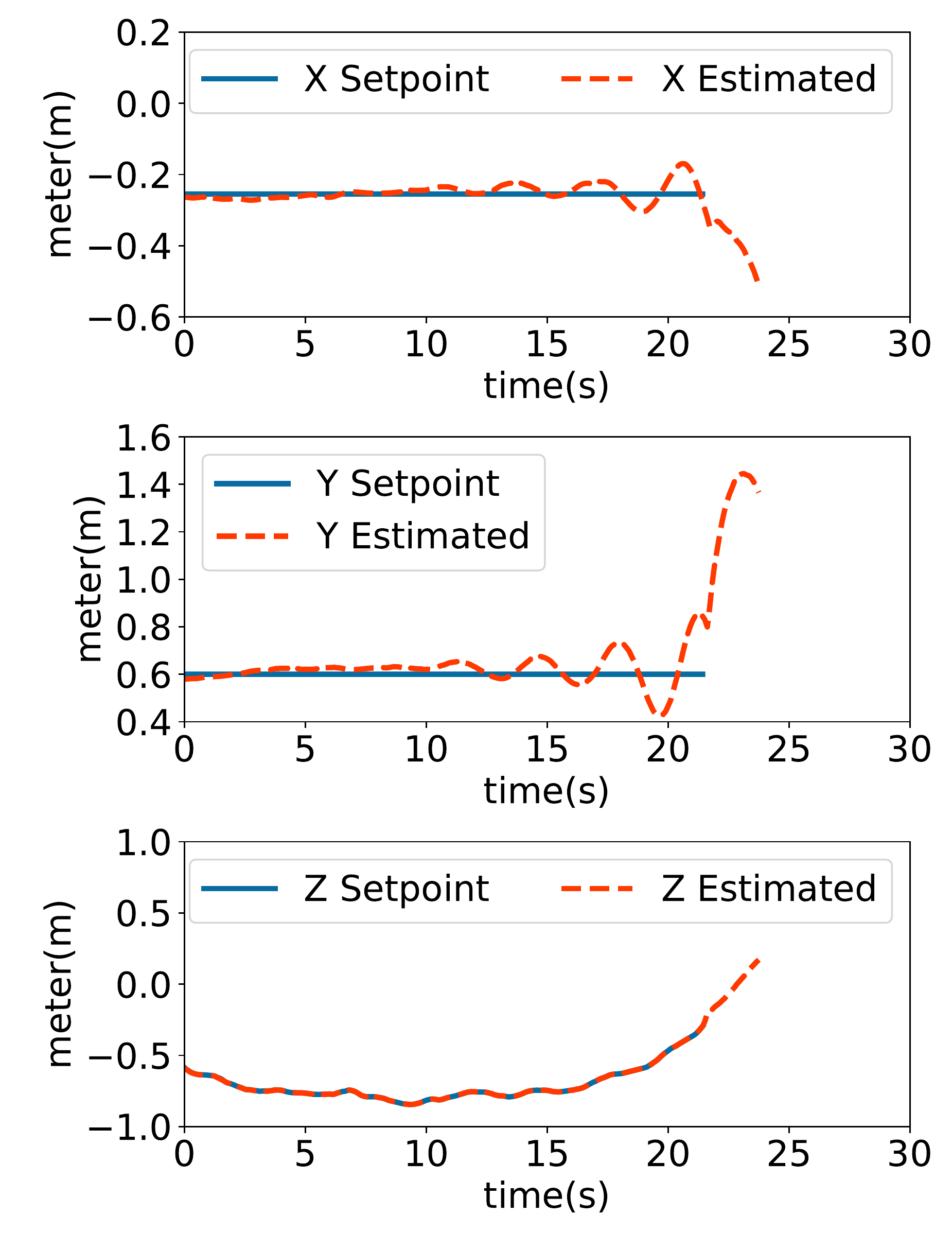}\vspace{-1em}
		\caption{Local position X, Y and Z for the quadcopter without MemGuard. The Bandwidth attack starts at 10 seconds and the drone crashes shortly after.}\label{f:withoutmem}
	\end{minipage}
		\begin{minipage}[t]{0.02\linewidth}
		\centering
\includegraphics[width=0.9\linewidth]{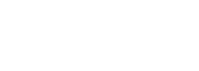}
	\end{minipage}
	\begin{minipage}[t]{0.23\linewidth}
		\centering
		\includegraphics[width=1.0\linewidth]{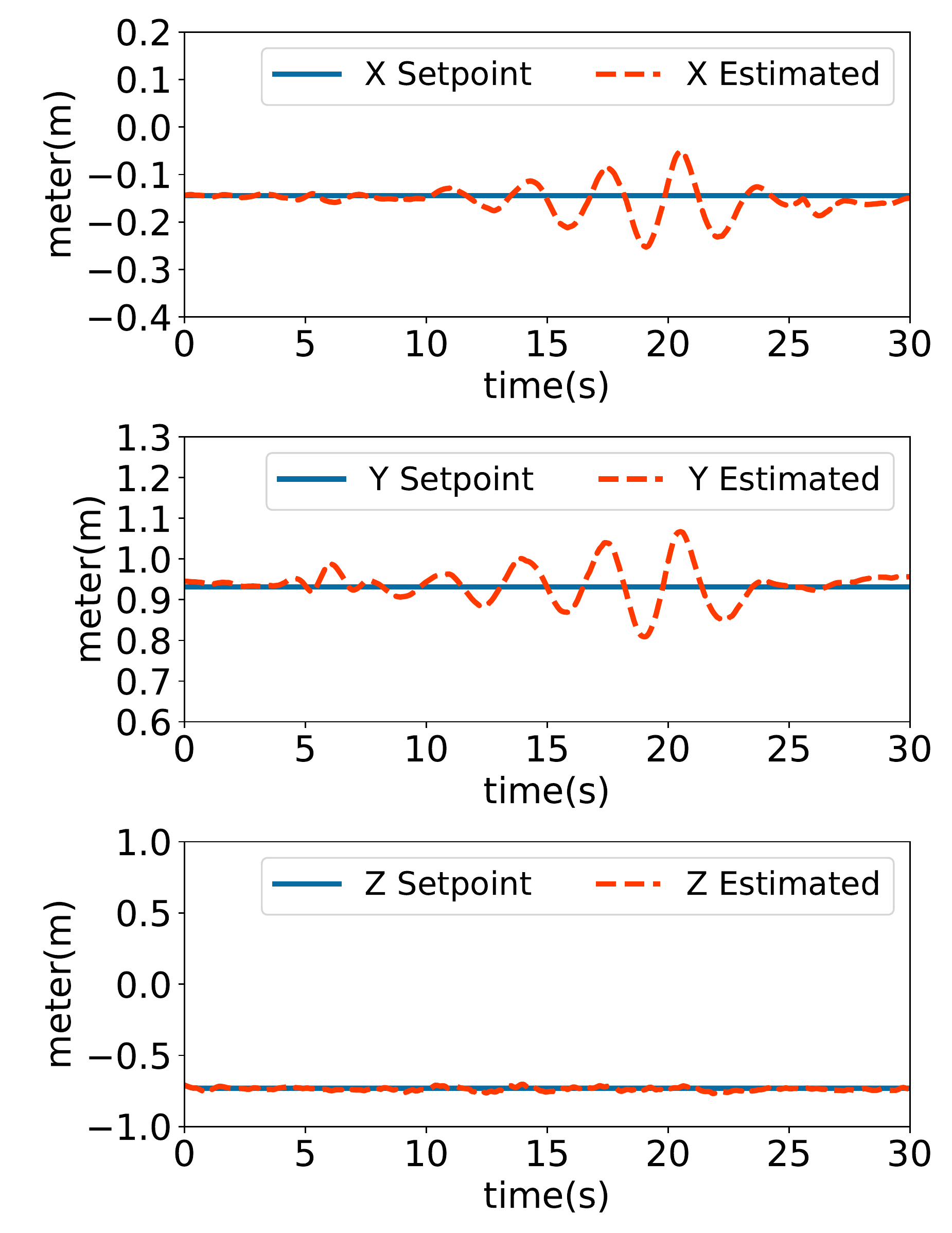}\vspace{-1em}
	\caption{Local position X, Y and Z for the quadcopter with MemGuard. The drone oscillates but is still stable.}\label{f:withmem}
	\end{minipage}
			\begin{minipage}[t]{0.01\linewidth}
		\centering
\includegraphics[width=0.9\linewidth]{fig/scape.pdf}
	\end{minipage}
	\begin{minipage}[t]{0.23\linewidth}
		\centering
		\includegraphics[width=1.0\linewidth]{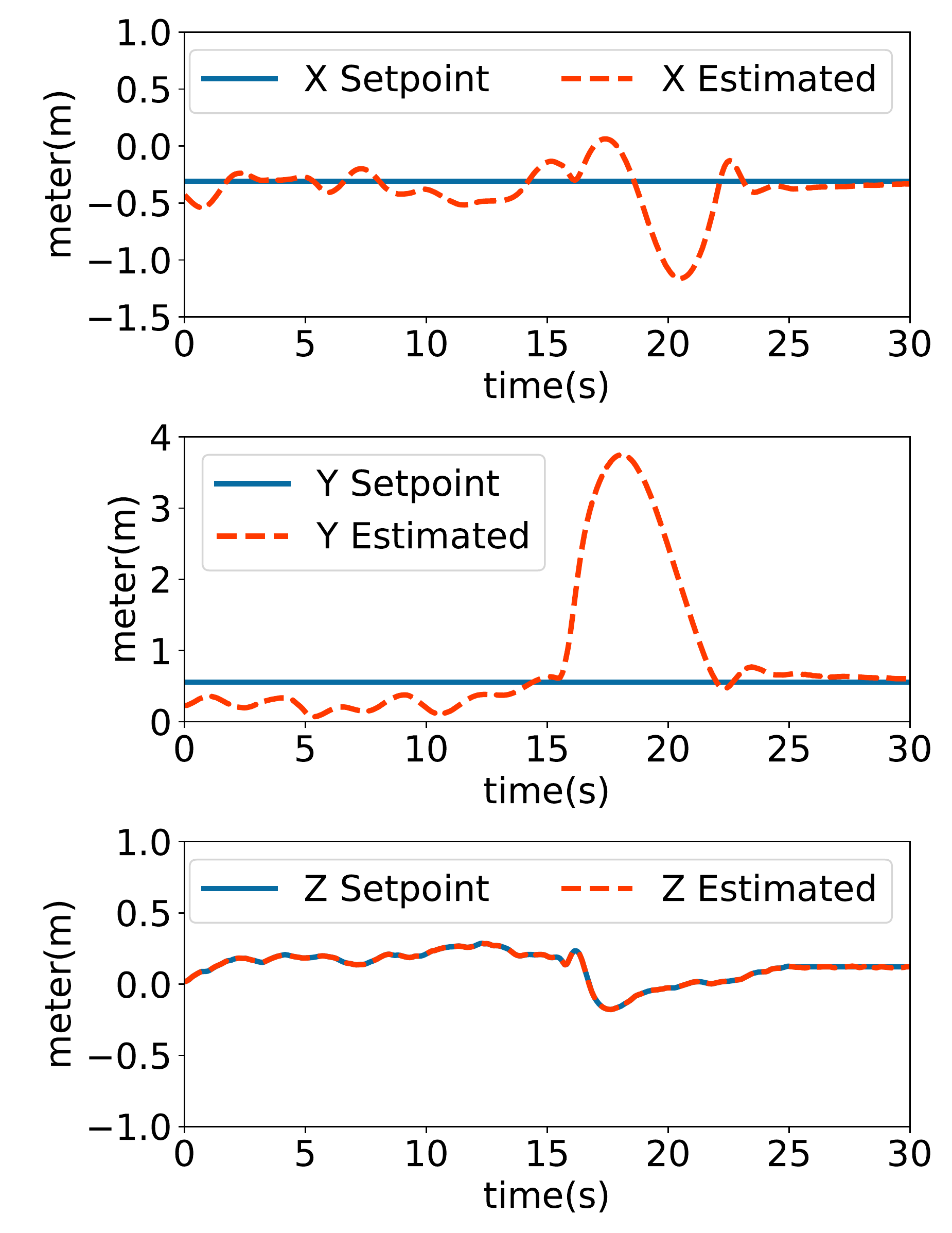}\vspace{-1em}
	\caption{Local position X, Y and Z for the quadcopter. The complex controller is killed at 12 seconds. The monitor switches the control back to the safety controller to stabilize the drone.}\label{f:killpx4}
	\end{minipage}
	\vspace{-1em}
\begin{minipage}[t]{0.01\linewidth}
		\centering
\includegraphics[width=0.9\linewidth]{fig/scape.pdf}
	\end{minipage}
	\begin{minipage}[t]{0.23\linewidth}
		\centering
		\includegraphics[width=1.0\linewidth]{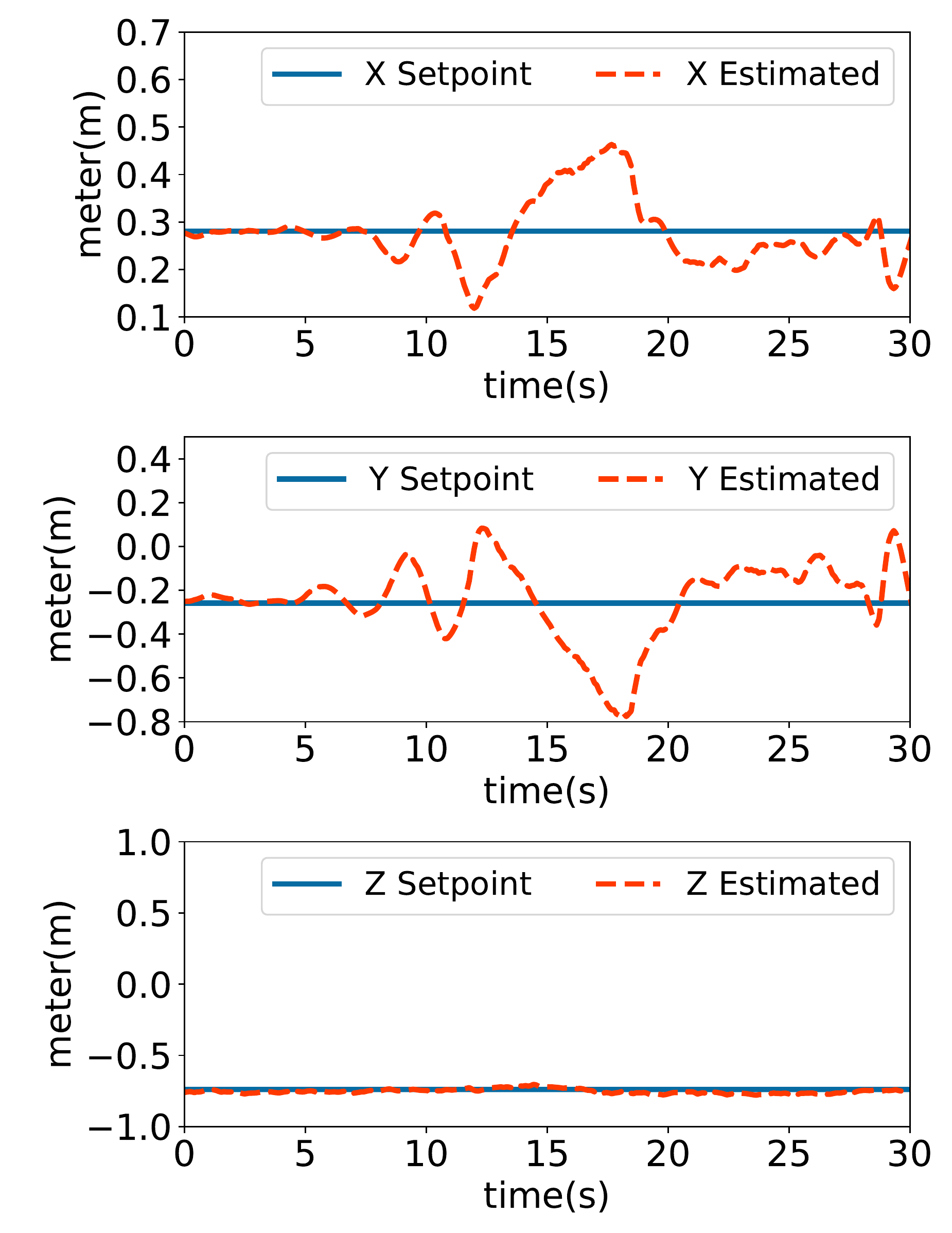}\vspace{-1em}
		\caption{Local position X, Y and Z for the quadcopter under UDP DoS attack. The drone is able to recover from unstable flight.}\label{f:udp}
	\end{minipage}
\end{figure*}
\subsection{Container and VM overhead comparison}
To show how lightweight the container implementation is, we compare the CPU idle rates when running only one container and virtual machine respectively on the host.
The container is configured as Section \ref{s:imple}.
The virtual machine is created with QEMU \cite{qemu} v3.0.0 and emulates an ARM Versatile/PB (ARM926EJ-S) platform with 256MB memory. 
In the virtual machine, we run the same Linux kernel image as the host.
As the results shown in Table \ref{t:systemoverheaad}, the overhead of running containers is closed to the native OS case and is much lower than running VMs.


\begin{table}[htbp]
\caption{System overhead comparison.} \label{t:systemoverheaad}
\begin{center}
\begin{tabular}{|l|c|c|c|c|}
\hline
\diagbox[width=10em]{Cases}{Overhead}{\#CPU} & CPU0 & CPU1 & CPU2 & CPU3 \\
\hline
No container nor VM& 0.95 & 0.99 & 0.99 & 0.99 \\
\hline
One VM & 0.86 & 0.83 & 0.81 & 0.77 \\
\hline
One container & 0.95 & 0.99 & 0.99 & 0.98  \\
\hline
\end{tabular}
\label{tab1}
\end{center}
\end{table}

\subsection{Memguard protect drone from memory DoS attack}
The attacker can run a memory intensive process inside the container to launch a memory DoS attack against the HCE. 
We used the Bandwidth from Isolbench, a benchmark that reads or writes a large array sequentially, to simulate the attacker's behavior. 
In this experiment, the Bandwidth task is the only process running inside the container, as this allows it to utilize all container resources exclusively and can maximize the potential damage. 
The attacker launches the Bandwidth task mid-fly and the performance of the drone is compared with MemGuard enabled and turned off. 
The MemGuard budget for the CCE is set to a value that allows the complex controller to run without problem. 

In the case without MemGuard, Figure \ref{f:withoutmem}, the drone starts to drift right after the Bandwidth task is launched by the attacker (at 15 seconds) and results in a crash shortly after. 
When the MemGuard is enabled, as in Figure \ref{f:withmem}, the drone oscillates for a short time but then managed to stabilize itself.  


\subsection{Security Monitoring defends UDP DoS attack}

The attacker may use the UDP channel to initiate a DoS attack against the HCE. 
To demonstrate this attack, we launched a program mid-fly that continuously send packets to the UDP port that the HCE is listening on. 

The results are presented in Figure \ref{f:udp}. After the program starts at 8 seconds, the drone starts circling and the radius gradually increases. 
Then attitude error control kicks in, killing the receiving thread on HCE and switching the control to safety controller, and brings the drone back to a stable state.

\subsection{Security Monitoring defends safety attack}

As the complex controller has potential vulnerabilities, the attacker might choose to kill it to not only damage the drone's safety but also maximize the resource used for attack. 
In this case, shown in Figure \ref{f:killpx4}, the attacker shutdown the complex controller while the drone is flying, at 12 seconds. 
The security monitor detects that the output from CCE has not been received for some time, then kills the receiving thread and switches to the output from the safety controller, shown in Figure \ref{f:killpx4}.


\section{Related work} \label{s:rela}


Container's lightweight footprint makes it a potential candidate for running real-time applications.  
Mao et al. explored using real-time container for Radio Access Networks (RANs) to minimize the latencies in cloud network\cite{mao2015minimizing}, and Masek studied the performance overhead of using containers for cyber-physical system application in \cite{masek2006container}. 
However, they used testing application to measure container latencies with a real-time patch kernel but not running the actual program to be used in the container. 


For DoS attacks in embedded systems, Wood et al. studied the defense mechanisms in each layer of a sensor network in \cite{wood2002denial} and Raymond et al. \cite{raymond2008denial} explored the defense mechanisms in wireless network.
These defense mechanisms focused more on the security of different network layers.
Chelladhurai et al. proposed security algorithms and methods to address DoS attacks related issues in the Docker container in \cite{chelladhurai2016securing}.
The goal is to protect the container from the attack coming from the network, instead of securing the host applications of the DoS attacks from the containers.

The Simplex Architecture \cite{sha2001using,seto1999case} has been used to provide guarantees for systems that use unreliable control.
The System-Level Simplex \cite{bak2009system} used hardware/software co-design to achieve fault tolerance in the system level. 
Based on this, Mohan et al. \cite{mohan2013s3a} used a combination of trusted hardware and OS techniques to enhance the security and safety of the real-time controller.  
The VirtualDrone \cite{yoon2017virtualdrone} used the virtual machine to implement the Simplex architecture and provides attack-resilience for a range of attacks.
However, the high latency introduced by the virtual machine makes it impossible to enforce more real-time resource control. 
The Simplex architecture has also been implemented using two heterogeneous hardware in \cite{vivekanandan2016simplex}.
However, they focus more on software bugs and transient hardware faults.

\section{Conclusion}\label{s:conclusion}

In this paper, we developed a DoS attack-resilient control framework, ContainerDrone, for real-time UAV system using containers. 
It is composed of the verified HCE, which runs a safety controller that implements only safety critical functionalities, and the CCE that runs a complex controller with better performance but is potentially unsafe.
The framework protects the HCE from the DoS attack launched inside the container by limiting attacker's access to three critical system resources: CPU, memory, and communication channel.
We implemented the framework on a prototype quadcopter using RPi3B and the open-source Docker container software.
Through experiments, we demonstrated the proposed framework can defend against various types of DoS attack launched from CCE effectively.
In the future, we plan to provide hard real-time proof and schedulability analysis for container drone and extend the framework to other types of unmanned vehicles.



\section*{Acknowledgment}

This project is sponsored in part by NSF 1739732 and by N00014-17-1-2783, and China Scholarship Council under Grant No.: 201706080092. The work was carried out at the Intelligent Robotics Laboratory, Coordinated Science Laboratory, University of Illinois at Urbana-Champaign.

\bibliographystyle{IEEEtran}
\bibliography{ref}

\color{red}

\end{document}